\documentclass[iop]{emulateapj}

\newcommand{\kms}{\ensuremath{\rm km\,s^{-1}}}
\newcommand{\ms}{\ensuremath{\rm m\,s^{-1}}}

\newcommand{\gcmc}{\ensuremath{\rm g\,cm^{-3}}}

\newcommand{\teff}{\ensuremath{T_{\rm eff}}}

\newcommand{\logg}{\ensuremath{\log{g}}}
\newcommand{\vsini}{\ensuremath{v \sin{i}}}
\newcommand{\feh}{\ensuremath{\rm [Fe/H]}}

\newcommand{\rsun}{\ensuremath{R_\sun}}
\newcommand{\msun}{\ensuremath{M_\sun}}
\newcommand{\lsun}{\ensuremath{L_\sun}}

\newcommand{\rstar}{\ensuremath{R_\star}}
\newcommand{\mstar}{\ensuremath{M_\star}}
\newcommand{\lstar}{\ensuremath{L_\star}}

\newcommand{\teffstar}{\ensuremath{T_{\rm eff\star}}}

\newcommand{\loggstar}{\ensuremath{\log{g_{\star}}}}

\newcommand{\rpl}{\ensuremath{R_{p}}}
\newcommand{\mpl}{\ensuremath{M_{p}}}

\newcommand{\rhopl}{\ensuremath{\rho_{p}}}

\newcommand{\rjup}{\ensuremath{R_{\rm J}}}
\newcommand{\mjup}{\ensuremath{M_{\rm J}}}

\newcommand{\wspitzer}{\emph{Warm-Spitzer}}
\newcommand{\spitzer}{\emph{Spitzer}}
\newcommand{\kepler}{\emph{Kepler}}

\newcommand{\kep}{Kepler-14}
\newcommand{\kepb}{Kepler-14b}
\newcommand{\kepcurLCP}{\ensuremath{6.7901230 \pm 0.0000043}}
\newcommand{\kepcurLCPshort}{6.790}
\newcommand{\kepcurLCT}{\ensuremath{2454971.08737 \pm0.00018}}
\newcommand{\kepcurDuration}{\ensuremath{0.2591^{+0.0040}_{-0.0039}}}
\newcommand{\kepcurDurationNoCorr}{\ensuremath{0.2561^{+0.0025}_{-0.0025}}}

\newcommand{\kepSeparation}{0.29\arcsec}
\newcommand{\kepKIC}{KIC~10264660}
\newcommand{\keptwomass}{2MASS~J19105011+4719589}
\newcommand{\kepRA}{\ensuremath{19^{\mathrm{h}}10^{\mathrm{m}}50\fs{12}}}
\newcommand{\kepDec}{\ensuremath{+47^{\circ}19'58\farcs{98}}}
\newcommand{\kepMag}{\ensuremath{12.128}}
\newcommand{\kepMagDiff}{\ensuremath{0.44\pm0.10}}

\newcommand{\koi}{KOI-98}

\newcommand{\kepcurLCar}{\ensuremath{8.213^{+0.578}_{-0.093}}}
\newcommand{\kepcurLCarNoCorr}{\ensuremath{7.472^{+0.481}_{-0.371}}}
\newcommand{\kepcurLCrprstar}{\ensuremath{0.0569 \pm 0.0013}}
\newcommand{\kepcurLCrprstarNoCorr}{\ensuremath{0.0448^{+0.0008}_{-0.0002}}}
\newcommand{\kepcurLCimp}{\ensuremath{0.00^{+0.41}_{-0.00}}}
\newcommand{\kepcurLCimpNoCorr}{\ensuremath{0.531^{+0.088}_{-0.078}}}
\newcommand{\kepcurLCi}{\ensuremath{90.0^{+0.0}_{-2.8}}}
\newcommand{\kepcurLCiNoCorr}{\ensuremath{85.92^{+0.80}_{-0.92}}}

\newcommand{\kepcurRVK}{\ensuremath{401.7 \pm7.1}}
\newcommand{\kepcurRVKcorr}{\ensuremath{682.9 \pm7.1}}

\newcommand{\kepcurRVrms}{\ensuremath{16.2}}

\newcommand{\kepcurRVmean}{\ensuremath{6.53 \pm 0.30 }}

\newcommand{\kepcurSMEteff}{\ensuremath{6395 \pm 60}}
\newcommand{\kepcurSMElogg}{\ensuremath{4.11 \pm 0.10}}
\newcommand{\kepcurSMEfeh}{\ensuremath{+0.12 \pm 0.06}}
\newcommand{\kepcurSMEvsin}{\ensuremath{7.9 \pm 1.0}}

\newcommand{\kepcurFIESteff}{\ensuremath{6378 \pm 80}}
\newcommand{\kepcurFIESlogg}{\ensuremath{4.02 \pm 0.21}}
\newcommand{\kepcurFIESfeh}{\ensuremath{0.00 \pm 0.12}}
\newcommand{\kepcurFIESvsin}{\ensuremath{10.6 \pm 1.0}}

\newcommand{\kepcurYYmlong}{\ensuremath{1.512 \pm 0.043}}
\newcommand{\kepcurYYrlong}{\ensuremath{2.048_{-0.084}^{+0.112}}}
\newcommand{\kepcurYYlogg}{\ensuremath{3.994_{-0.036}^{+0.028}}}
\newcommand{\kepcurYYlum}{\ensuremath{6.29_{-0.58}^{+0.75}}}
\newcommand{\kepcurYYage}{\ensuremath{2.2_{-0.1}^{+0.2}}}

\newcommand{\kepcurYYmlongNoCorr}{\ensuremath{1.604^{+0.057}_{-0.060}}}
\newcommand{\kepcurYYrlongNoCorr}{\ensuremath{2.358_{-0.166}^{+0.147}}}
\newcommand{\kepcurYYloggNoCorr}{\ensuremath{3.899_{-0.041}^{+0.049}}}
\newcommand{\kepcurYYlumNoCorr}{\ensuremath{8.30_{-1.17}^{+1.16}}}
\newcommand{\kepcurYYageNoCorr}{\ensuremath{2.0_{-0.2}^{+0.2}}}

\newcommand{\kepcurXdist}{\ensuremath{980}}
\newcommand{\kepcurPPmNoCorr}{\ensuremath{5.14^{+0.15}_{-0.16}}}
\newcommand{\kepcurPPrNoCorr}{\ensuremath{1.036_{-0.084}^{+0.075}}}
\newcommand{\kepcurPPm}{\ensuremath{8.40^{+0.19}_{-0.18}}}
\newcommand{\kepcurPPr}{\ensuremath{1.136_{-0.054}^{+0.073}}}
\newcommand{\kepcurPPrho}{\ensuremath{7.1\pm1.1}}
\newcommand{\kepcurPPrhoNoCorr}{\ensuremath{5.7_{-1.0}^{+1.5}}}

\shorttitle{\kepb}
\shortauthors{Buchhave et al.}

\begin{document}

\title{\kepb: A massive hot Jupiter transiting an F star in a close visual binary}

\altaffiltext{1}{Niels Bohr Institute, University of Copenhagen, DK-2100 Copenhagen, Denmark}
\altaffiltext{2}{StarPlan, University of Copenhagen, Denmark}
\altaffiltext{3}{Harvard-Smithsonian Center for Astrophysics, Cambridge, MA 02138, USA}
\altaffiltext{4}{Hubble Fellow}
\altaffiltext{5}{NASA Ames Research Center, Moffett Field, CA 94035, USA}
\altaffiltext{6}{NASA Exoplanet Science Institute/California Institute of Technology, Pasadena, CA 91109, USA}
\altaffiltext{7}{Lunar and Planetary Laboratory, University of Arizona, Tucson, AZ 85721, USA}
\altaffiltext{8}{Yale University, New Haven, CT 06520, USA}
\altaffiltext{9}{Jet Propulsion Laboratory/California Institute of Technology, Pasadena, CA 91109, USA}
\altaffiltext{10}{Space Telescope Science Institute, Baltimore, MD 21218, USA}
\altaffiltext{11}{University of California, Berkeley, Berkeley, CA 94720, USA}
\altaffiltext{12}{SETI Institute/NASA Ames Research Center, Moffett Field, CA 94035, USA}
\altaffiltext{13}{Steward Observatory, Uni. of Arizona, Tucson, AZ, USA}
\altaffiltext{14}{San Jose State University, San Jose, CA 95192, USA}
\altaffiltext{15}{Las Cumbres Observatory, Goleta, CA 93117, USA}
\altaffiltext{16}{University of Texas, Austin, TX 78712, USA}
\altaffiltext{17}{Solar System Exploration Division, NASA Goddard Space Flight Center, Greenbelt, MD 20771, USA}
\altaffiltext{18}{Lowell Observatory, Flagstaff, AZ 86001, USA}
\altaffiltext{19}{National Optical Astronomy Observatory, Tucson, AZ 85719, USA}
\altaffiltext{20}{University of Florida, Gainesville, FL 32611}
\altaffiltext{21}{Department of Astronomy and Astrophysics, University of California, Santa Cruz, CA 95064, USA}
\altaffiltext{22}{Orbital Sciences Corporation/NASA Ames Research Center, Moffett Field, CA 94035}
\altaffiltext{23}{Department of Physics, Southern Connecticut State University, New Haven, CT 06515, USA}
\altaffiltext{24}{Massachusetts Institute of Technology, Cambridge, MA 02159, USA}
\altaffiltext{25}{BAER Institute, Moffett Field, CA 94035, USA}
\altaffiltext{26}{MSFC, Huntsville, AL 35805 USA}

\author{
Lars~A.~Buchhave\altaffilmark{1,2},
David~W.~Latham\altaffilmark{3},
Joshua~A.~Carter\altaffilmark{3,4},
Jean-Michel~D\'esert\altaffilmark{3},
Guillermo~Torres\altaffilmark{3},
Elisabeth~R.~Adams\altaffilmark{3},
Stephen~T.~Bryson\altaffilmark{5},
David~B.~Charbonneau\altaffilmark{3},
David~R.~Ciardi\altaffilmark{6},
Craig~Kulesa\altaffilmark{7},
Andrea~K.~Dupree\altaffilmark{3},
Debra~A.~Fischer\altaffilmark{8},
Fran\c{c}ois~Fressin\altaffilmark{3},
Thomas~N.~Gautier~III\altaffilmark{9},
Ronald~L.~Gilliland\altaffilmark{10},
Steve~B.~Howell\altaffilmark{5},
Howard~Isaacson\altaffilmark{11},
Jon~M.~Jenkins\altaffilmark{12},
Geoffrey~W.~Marcy\altaffilmark{11},
Donald~W.~McCarthy\altaffilmark{13},
Jason~F.~Rowe\altaffilmark{12},
Natalie~M.~Batalha\altaffilmark{14},
William~J.~Borucki\altaffilmark{5},
Timothy~M.~Brown\altaffilmark{15},
Douglas~A.~Caldwell\altaffilmark{12},
Jessie~L.~Christiansen\altaffilmark{12},
William~D.~Cochran\altaffilmark{16},
Drake~Deming\altaffilmark{17},
Edward~W.~Dunham\altaffilmark{18},
Mark~Everett\altaffilmark{19},
Eric~B.~Ford\altaffilmark{20},
Jonathan~J.~Fortney\altaffilmark{21},
John~C.~Geary\altaffilmark{3},
Forrest~R.~Girouard\altaffilmark{22},
Michael~R.~Haas\altaffilmark{5},
Matthew~J.~Holman\altaffilmark{3},
Elliott~Horch\altaffilmark{23},
Todd~C.~Klaus\altaffilmark{22},
Heather~A.~Knutson\altaffilmark{11},
David~G.~Koch\altaffilmark{5},
Jeffrey~Kolodziejczak\altaffilmark{26},
Jack~J.~Lissauer\altaffilmark{5},
Pavel~Machalek\altaffilmark{12},
Fergal~Mullally\altaffilmark{12},
Martin~D.~Still\altaffilmark{25},
Samuel~N.~Quinn\altaffilmark{3},
Sara~Seager\altaffilmark{24}
Susan~E.~Thompson\altaffilmark{12},
Jeffrey~Van~Cleve\altaffilmark{12}
}

\begin{abstract}
We present the discovery of a hot Jupiter transiting an F star in a close visual ($0.3 \arcsec$ sky projected angular separation) binary system. The dilution of  the host star's light by the nearly equal magnitude stellar companion ($\sim 0.5$ magnitudes fainter) significantly affects the derived planetary parameters, and if left uncorrected, leads to an underestimate of the radius and mass of the planet by 10\% and 60\%, respectively. Other published exoplanets, which have not been observed with high-resolution imaging, could similarly have unresolved stellar companions and thus have incorrectly derived planetary parameters. \kepb\ (\koi) has a period of $P=\kepcurLCPshort$ days and correcting for the dilution, has a mass of $\mpl = \kepcurPPm\,\mjup$ and a radius of $\rpl = \kepcurPPr\,\rjup$, yielding a mean density of $\rhopl = \kepcurPPrho\,\gcmc$. 
\end{abstract}

\section{Introduction}
\label{intro_sect}
Kepler is a space-based telescope using transit photometry to determine the frequency and characteristics of planets and planetary systems \citep{Borucki2010}. The instrument was launched in March 2009 and is a wide field-of-view (115 square degrees) photometer comprised of a 0.95-meter effective aperture Schmidt telescope that monitors the brightness of about 150,000 stars. Recently, the first four months of photometric data were released and over 1200 transiting planet candidates were identified \citep{Borucki2011}. \kepb\ was identified among the 1235 candidates as \koi. Because of its short period ($P=\kepcurLCPshort$ days) and its relatively deep transit signal, it was  identified very early in the mission and quickly passed on to the Kepler Follow-up Program (KFOP) for further investigation.

\kepb\ was scrutinized for evidence of astrophysical false positives and survived the initial follow-up stage. We therefore gathered high resolution, high signal-to-noise ratio (SNR) spectra in order to extract precise radial velocities (RVs) using the FIber-fed \'Echelle Spectrograph (FIES) on the Nordic Optical Telescope (NOT) at La Palma in October 2009. The observations yielded a spectroscopic orbit in phase with the photometric observations from Kepler. Meanwhile, other ground based observations were gathered and speckle imaging revealed that the host star of \kepb\ was in fact not a single star, but a nearly equal magnitude binary system with a sky projected separation of only \kepSeparation. This complicated the analysis of the system, and \kepb\ was therefore not readied for publication together with the initial batch of five  Kepler planets.

In this paper, we present the confirmation of \kepb\ as a planetary companion and analyze the photometric and spectroscopic data taking into account the effect of the dilution by the stellar companion on the derived planetary parameters. Without the high spatial resolution imaging, the stellar companion would not have been detected and \kep\ would have been published with incorrect planetary parameters. Other published transiting planets that have not been observed with high spatial resolution imaging might suffer the same problem.

\section{Kepler photometry}
\label{sec:keplerPhoto}
\begin{figure}
\centering
\plotone{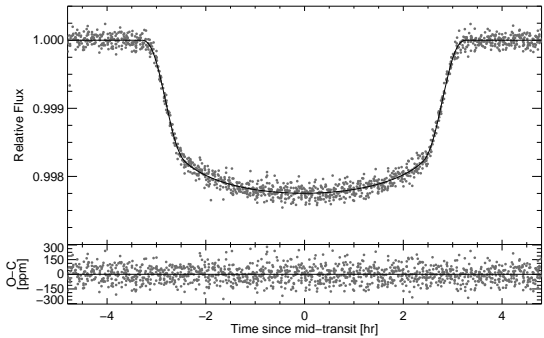}
\caption{Light curve for \kepb. The upper panel shows the photometry folded with the orbital period of \kepcurLCPshort\ days. The fitted transit model is overplotted as a solid line. Residuals from the fit are displayed in the bottom panel.}
\label{fig:kepler_pho}
\end{figure}

We use the long cadence photometry (29.4 minute accumulations) of \kep\ (\kepKIC, \kepRA, \kepDec, J2000, KIC $r=\kepMag$~mag) obtained between 5 May 2009 to 7 September 2010 (Q0 through Q6) \citep{Jenkins2010a}. The photometry is processed in an analysis pipeline \citep{Jenkins2010b} to produce corrected pixel data. Simple aperture photometry sums are formed to produce a photometric time series, which was detrended as explained in \citet{Koch2010}. The photometric data folded with the orbital period of \kepcurLCPshort\ days are shown in Figure \ref{fig:kepler_pho}. The numerical data are available electronically from the Multi Mission Archive at the Space Telescope Science Institute (MAST) Web site\footnote{http://archive.stsci.edu/kepler/}.

\section{Follow-up observations}
\label{sec:followupObs}

\subsection{High-resolution speckle imaging}
\label{sec:speckle}
\begin{deluxetable}{lrrr}
\tablewidth{0pc}
\tablecaption{
Speckle measurements of \kep{}
\label{tab:speckle}
}
\tablehead{
\colhead{Band}                        		   &
\colhead{Separation\tablenotemark{a}}                           &
\colhead{Position angle\tablenotemark{b}}                       &
\colhead{$\Delta_{mag}$}                       
}
\startdata
V & $0.286\pm0.04$ &  $143.67\pm0.07$ & $0.52\pm0.05$\\ 
R & $0.289\pm0.01$ &  $143.67\pm0.07$ & $0.54\pm0.12$\\ 
I & $0.289\pm0.02$ &  $143.91\pm0.05$ & $0.45\pm0.04$\\ 
[-1.5ex]
\enddata
\tablenotetext{a}{
In arcseconds.
}
\tablenotetext{b}{
In degrees.
}
\end{deluxetable}

Speckle observations of \kep\ were obtained on 6 different nights between June and Oct 2010. The observations were obtained with the dual channel WIYN  telescope speckle camera recently described in \citet{Horch2011}. The data collection, reduction, and image reconstruction process is described in the aforementioned paper as well as in \citet{Howell2011}. The latter presents details of the 2010 season of observations for the Kepler follow-up program.
 
\begin{figure}
\centering
\plotone{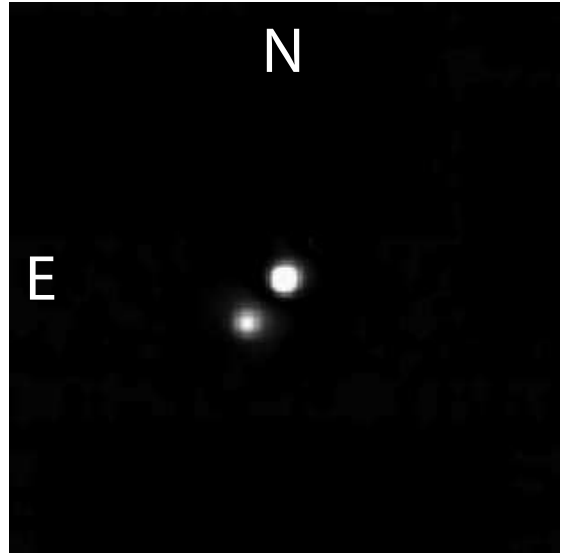}
\caption{Speckle image of \kep\ with its obvious companion star separated by 0.3\arcsec of nearly equal brightness ($\sim 0.5$ magnitude fainter).}
\label{fig:speckle}
\end{figure}

A spatially close (0.3\arcsec), nearly equal brightness ($\sim 0.5$ magnitude fainter) companion star was easily noted in the reconstructed speckle images. Table \ref{tab:speckle} gives the weighted mean values for the separation, position angle, and magnitude difference for our six speckle observations. The observations are weighted by the native seeing during the time of the speckle data collection as determined by the data reduction routines when fitting known single point source speckle standard stars obtained near in time to the Kepler star observations. Figure \ref{fig:speckle} shows one of the reconstructed images of \kep\ with its obvious companion star.

\begin{figure}
\centering
\plotone{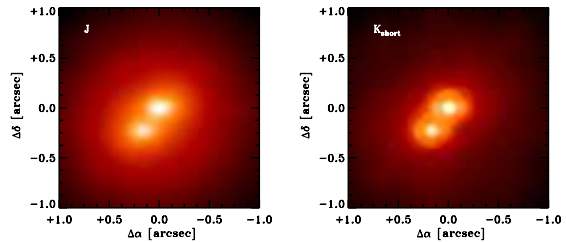}
\caption{$J$ and $K_{s}$ Palomar adaptive optics images of \kep, showing a $2\arcsec \times 2\arcsec$ field of view centered on the brighter star.}
\label{fig:AO_palo}
\end{figure}

\begin{figure}
\centering
\plotone{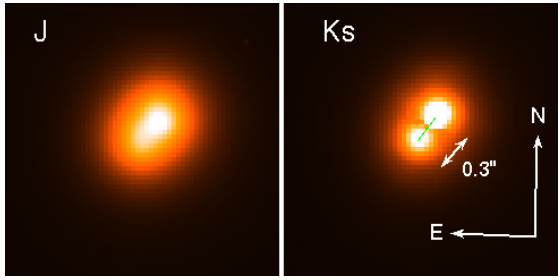}
\caption{MMT/ARIES AO images of \kep\ showing a $2.4\arcsec \times 2.4\arcsec$ field of view in $J$ and $K_s$ band. The binary nature of \kep\ is clear from the figure, with the fainter star, B, offset by $0.29 \pm 0.01\arcsec$ to the south-east.}
\label{fig:AO_ARIES}
\end{figure}

\subsection{High-resolution Palomar AO imaging}
\label{sec:AO_palomar}
Near-infrared adaptive optics imaging of \kep\ was obtained on the night of 03 July 2010 UT with the Palomar Hale 200-inch telescope and the PHARO near-infrared camera \citep{Hayward2001} behind the Palomar adaptive optics system \citep{Troy2000}.  PHARO, a $1024\times1024$ HgCdTe infrared array, was utilized in 25.1 mas/pixel mode yielding a field of view of $25\arcsec$.  Observations were performed in both the $J$ ($\lambda_0 = 1.25\,\mu$m) and $K_s$ ($\lambda_0 = 2.145\,\mu$m) filters.  The data were collected in a standard 5-point quincunx dither pattern of $5\arcsec$ steps interlaced with an off-source ($60\arcsec$ East) sky dither pattern. The integration time per source was 24.5 seconds at $J$ and 12.75 seconds at $K_s$.  A total of 20 frames were acquired at $J$ and $K_s$ for a total on-source integration time of 8.5 minutes and 4 minutes, respectively.  The individual frames were reduced with a set of custom IDL routines written for the PHARO camera and were combined into a single final image.  The adaptive optics system guided on the primary target itself; the widths of the central cores of the resulting point spread functions were $FWHM = 0.15\arcsec$ at $J$ and $FWHM = 0.1\arcsec$ at $K_s$.  The final co-added images at $J$ and $K_s$ are shown in Figure~\ref{fig:AO_palo}.

Other than the nearby object, two other sources were detected within $10\arcsec$ of the primary target.  The second closest object is separated from \kep\ by $5.5\arcsec$ to the northwest and has infrared magnitudes of $J = 18.08\pm0.04$ and $K_s = 17.28\pm0.01$.  An additional source was detected to the southwest at a distance of 6.2$\arcsec$ having infrared magnitudes of $J = 19.00\pm0.05$ and $K_s = 18.14\pm0.04$. 

The close pair was easily resolved by the adaptive optics at both $J$ and $K_s$. The pair is separated by $\sim 0.28\arcsec$ with a position angle of $142^\circ$ east of north. The pair has magnitude differences of $\Delta J = 0.34 \pm 0.01$ and $\Delta Ks = 0.40 \pm 0.01$. The brighter infrared (and optical) source (component A) is the northwestern star. 
\subsection{High-resolution ARIES AO imaging}
\label{sec:AO_aries}
High-resolution AO images of \kep\ were obtained using the ARIES instrument on the 6.5-m MMT. ARIES is a near-infrared diffraction-limited imager and spectrograph. On 8 November, 2009 it was operated in the f/15 mode, with a 40\arcsec x40\arcsec field of view and a pixel scale of 0.04\arcsec/pixel. All images of \kep\ had exposure times of 10 seconds, with 16 images in $J$ (in a 4-point, 4\arcsec dither pattern) and 19 images taken in $K_s$ (16 in a 4-point, 4\arcsec dither pattern, and 3 images at other offsets). The images for each filter were calibrated using standard IRAF\footnote{IRAF is distributed by the National Optical Astronomy Observatories, which are operated by the Association of Universities for Research in Astronomy, Inc., under cooperative agreement with the National Science Foundation.} procedures, and combined and sky-subtracted using the IRAF task {\it xdimsum}.

In both $J$ and $K_s$, the binary appearance of \kep\ is clear, with the fainter star, B, offset by $0.29 \pm 0.01\arcsec$ to the south-east. The separation is comparable to the image FWHM (0.5\arcsec in $J$ and 0.3\arcsec in $K_s$). The relative magnitudes were estimated by PSF fitting, yielding  $\Delta J = 0.398 \pm 0.008$ and $\Delta K = 0.490 \pm 0.005$.

The delta magnitudes from the Palomar and ARIES AO imaging are thus similar, but the difference of 0.06 magnitude in $J$ and 0.09 magnitude in $K_s$ suggests that the accuracy is worse than implied by the formal precision. We combined all the Speckle and AO imaging results for the assessment of the dilution from the nearby companion.

\subsection{High-resolution high SNR spectroscopy}
\label{sec:spectroscopy}

\begin{deluxetable}{rrrrrr}
\tablewidth{0pc}
\tablecaption{
Relative Radial-Velocity Measurements of \kep{}
\label{tab:rvs}
}
\tablehead{
\colhead{HJD}                           &
\colhead{Phase}                         &
\colhead{RV}                            &
\colhead{\ensuremath{\sigma_{\rm RV}}}  &
\colhead{BS}                            &
\colhead{\ensuremath{\sigma_{\rm BS}}}  \\
\colhead{(days)}                        &
\colhead{(cycles)}                      &
\colhead{$(\ms)$}                         &
\colhead{$(\ms)$}                         &
\colhead{$(\ms)$}                         &
\colhead{$(\ms)$}
}
\startdata
2455048.454298 &  11.394 &$   -246.1 $&$ 25.0 $&$  21.4 $&$  7.4  $\\ 
2455052.427895 &  11.979 &$     31.5 $&$ 17.0 $&$   4.2 $&$  6.5  $\\ 
2455107.428247 &  20.079 &$   -219.1 $&$ 14.2 $&$  -6.8 $&$  4.9  $\\ 
2455108.417046 &  20.225 &$   -385.2 $&$ 15.4 $&$  -4.0 $&$  5.4  $\\ 
2455109.356436 &  20.363 &$   -285.7 $&$ 20.3 $&$  15.0 $&$ 10.8  $\\ 
2455109.415242 &  20.372 &$   -269.0 $&$ 20.7 $&$  -1.7 $&$  5.3  $\\ 
2455111.453147 &  20.672 &$    349.5 $&$ 14.2 $&$ -30.5 $&$  6.9  $\\ 
2455112.462361 &  20.821 &$    393.2 $&$ 18.0 $&$ -41.2 $&$  8.7  $\\ 
2455113.456678 &  20.967 &$     89.2 $&$ 19.6 $&$  -5.8 $&$  7.0  $\\ 
2455114.494316 &  21.120 &$   -260.0 $&$ 25.6 $&$  18.4 $&$  6.3  $\\ 
2455115.492291 &  21.267 &$   -404.0 $&$ 30.2 $&$  51.2 $&$ 11.0  $\\ 
2455122.475354 &  22.295 &$   -387.8 $&$ 20.3 $&$  -0.3 $&$ 11.2  $\\ 
2455123.417485 &  22.434 &$   -152.6 $&$ 19.4 $&$   9.4 $&$  7.3  $\\ 
2455125.407529 &  22.727 &$    398.5 $&$ 19.4 $&$ -29.4 $&$  7.3  $\\
[-1.5ex]
\enddata
\end{deluxetable}

\begin{figure}
\centering
\plotone{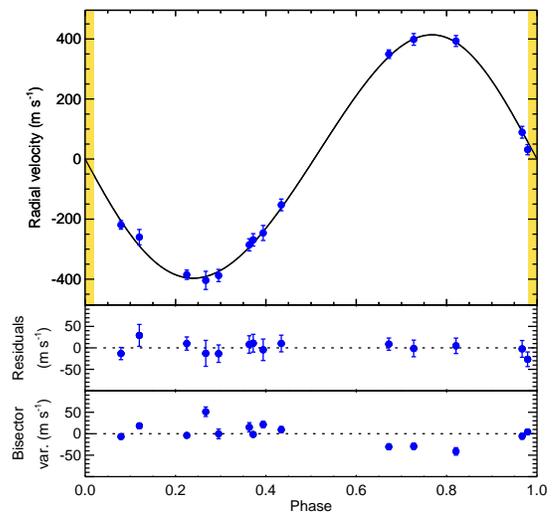}
\caption{Upper panel: Radial velocity measurements from the FIbre--fed \'Echelle Spectrograph (FIES) at the 2.5\,m Nordic Optical Telescope (NOT) at La Palma as a function of orbital phase with the best orbital fit overplotted. The velocity of the system has been subtracted and the fit assumes a circular orbit, fixing the ephemeris to that found by the photometry. Middle panel: Phased residuals of the velocities after subtracting the best fit model. The rms variation of the residuals is $\kepcurRVrms\ \ms$. Bottom panel: Variations of the bisector spans from the FIES spectra, with the mean value subtracted.}
\label{fig:rv_orbit}
\end{figure}

Spectroscopic observations of \kep\ were obtained using the FIbre--fed \'Echelle Spectrograph (FIES) at the 2.5\,m Nordic Optical Telescope (NOT) at La Palma, Spain \citep{Djupvik2010} as well as HIRES \citep{Vogt1994} mounted on the Keck~I telescope on Mauna Kea, Hawaii. We acquired 17 FIES spectra between 4 August and 20 October 2009, 3 of which were not used in the analysis because of very low SNR due to poor observing conditions. One HIRES template spectrum was also observed on 10 September 2009 and used to derive stellar parameters.

For HIRES, we set the spectrometer slit to $0\farcs86$, resulting in a resolving power of $\lambda/\Delta\lambda \approx 55,\!000$ with a wavelength coverage of
$\sim$3800--8000\,\AA\@. We reduced the HIRES spectrum following a procedure based on that described by \cite{Butler1996}.

For FIES, we used the medium and the high--resolution fibers ($1\farcs3$ projected diameter) with resolving powers of $\lambda/\Delta\lambda \approx 46,\!000$ and $67,\!000$, respectively, giving a wavelength coverage of $\sim 3600-7400$\,\AA\@. We used the wavelength range from approximately $\sim 3900-5800$\,\AA\@ to determine the radial velocities. The exposure time was approximately 60 minutes yielding a SNR from 20 to 65 per pixel (SNR of 38 to 120 per resolution element) over the wavelength range used. The rather large range in SNR is due to the variation in instrumental throughput and the stellar flux as a function of wavelength, and the lower throughput of the high resolution fiber.

The FIES spectra were rectified and cross correlated using a custom-built pipeline designed to provide precise radial velocities for \'Echelle spectrographs. The procedures are described in more detail in \citet{Buchhave2010}. The science exposures were bracketed by two thorium argon (ThAr) calibration images taken through the same fiber and extracted using the same pipeline as the science exposures. The ThAr images were then combined to form the basis for the fiducial wavelength calibration. Once the spectra had been extracted, a cross correlation was performed order by order using the strongest exposure as the template. The orders were cross correlated using a Fast Fourier Transform (FFT) and the cross correlation functions (CCFs) for all the orders were co-added and fitted with a Gaussian function to determine the radial velocity. Uncertainties of the individual velocities were estimated by $\sigma=RMS(v)/\sqrt{N}$, where $v$ is the radial velocity of the individual orders and $N$ is the number of orders.

The light from the fainter, but nearly equal magnitude stellar companion (B) dilutes the light of the brighter star (A). In Section \ref{sec:centroid}, use the photometric centroid to determine that it is the brighter star (A) which is undergoing transit and thus is the planet hosting star. The very small angular separation of \kepSeparation makes it impossible to separate the two stars on the fiber for the spectroscopic observations and it is thus necessary to account for the effect of the dilution on the measured radial velocities (see Section \ref{sec:specParam} and \ref{sec:RV_dilution}).

In the observed FIES and HIRES spectra, we did {\it not} see a composite spectrum in any of the observations. We would easily have been able to identify two cross correlation peaks from a composite spectrum, if the two stars did not have nearly equal radial velocities. The combination of the small angular separation of the two stars and the similar radial velocity makes the probability of the two stars being a chance alignment highly unlikely, and we therefore conclude that the two stellar components are gravitationally bound in a wide orbit yielding an undetectable radial velocity offset between the two spectra.

The radial velocity measurements of the combined light of the two components in \kep\ are reported in Table \ref{tab:rvs}. The radial velocities are relative, since they are measured relative to the strongest of the observed spectra adopted as the template. We made a separate estimate of the systemic velocity (the $\gamma$ velocity) by correlating the observed spectra against the synthetic library spectrum best matching the stellar parameters. We took the mean of these velocities and subtracted the gravitational redshift of the Sun ($0.636\,\kms$), which is not included in the calculation of the synthetic library spectra. We found the mean $\gamma$ velocity of \kepb\ to be $\gamma = \kepcurRVmean\ \kms$.

We fitted a circular orbit to the radial velocities reported in Table~\ref{tab:rvs}, adopting the photometric ephemeris, leaving only the orbital semi-amplitude, $K$, and an arbitrary RV offset as free parameters.  A plot of the orbital solution is shown in the top panel in Figure~\ref{fig:rv_orbit} with the residuals to the fit shown in the middle panel. The orbital parameters are listed in Table~\ref{tab:parameters}. Allowing the eccentricity to be a free parameter only reduced the velocity residuals by a small amount and yielded an eccentricity that was insignificant ($e = 0.035 \pm 0.020$). However, we included the eccentricity in the light curve analysis in Section \ref{sec:lightCurveAnalysis} mainly to allow for more realistic uncertainty estimates of the planetary parameters.

\subsection{Bisector analysis}
We carried out a bisector span analysis \citep{Queloz2001,Torres2007} of the FIES spectra to explore the possibility that the transit like events are due to an eclipsing binary blended with light from a third star. The bisector spans are plotted in the bottom panel of Figure \ref{fig:rv_orbit}.

Since the observed spectrum is a composite of the planet hosting star and its fainter companion, which we assume is stationary in velocity with respect to the reflex motion induced by the planet, we expect to see a slight asymmetry in the cross correlation peak as a function of phase. The bisector span is defined as the velocity of the bisector measured near the bottom of the CCF minus the velocity measured near the top, and we would thus expect a positive bisector span when the host star is moving toward us and a negative bisector span when the host star is moving away. In the bottom planet in Figure \ref{fig:rv_orbit}, we see a slight hint of this effect, with the bisector span being predominantly positive around phase 0.25 and predominantly negative around phase 0.75. The amplitude of the bisector spans is significantly less than the radial velocity semi-amplitude and the hint of variation is in the expected direction, which supports the interpretation that the radial velocity variations are due to a planetary companion.

\section{Warm-Spitzer Observations}
\label{sec:spitzer}

\kep\ was observed during one transit with \wspitzer/IRAC \citep{Werner2004,Fazio2004} at 4.5~\micron\ (program ID 60028). The observation occurred on UT 2010 August 07 and the visit lasted approximately 14~h 20~min. The data were gathered in full-frame mode ($256\times256$ pixels) with an exposure time of 30~s per image, which yielded 1700 images.

\begin{figure}
\centering
\plotone{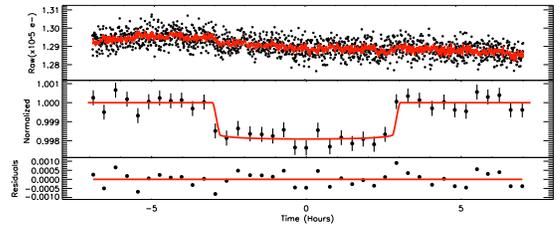}
\caption{\wspitzer\ transit light-curve of \kep\ observed in the IRAC band-pass at 4.5~\micron. Top panel: raw and unbinned transit light-curve. The red solid lines correspond to the best fit models which include the time and position instrumental decorrelations as well as the model for the planetary transit (see details in Section~\ref{sec:spitzer}). Middle panel: corrected, binned by 25~minutes and normalized transit light-curve with the best fit in red. Bottom panel: residuals of the data from the best fit.}
\label{fig:spitzerlightcurves}
\end{figure}

The method we used to produce photometric time series from the images is described by \cite{Desert2009}. It consists of finding the centroid position of the stellar point spread function (PSF) and performing aperture photometry using a circular aperture  on individual exposures. The images used are the Basic Calibrated Data (BCD) delivered by the \emph{Spitzer} archive. These files are corrected for dark current, flat-fielding, detector non-linearity and converted into flux units. We convert the pixel intensities to electrons using the information on the detector gain and exposure time provided in the FITS headers. This facilitates the evaluation of the photometric errors. We extract the UTC-based Julian date for each image from the FITS header (keyword DATE\_OBS) and correct to mid-exposure. We convert to TDB-based BJD using the \texttt{UTC2BJD}\footnote{{\tt http://astroutils.astronomy.ohio-state.edu/time/}} procedure developed by \citet{Eastman2010}. This program uses the JPL Horizons ephemeris to estimate the position of \emph{Spitzer Space Telescope} during the observations. We then correct for transient pixels in each individual image using a 20-point sliding median filter of the pixel intensity versus time. To do so, we compare each pixel's intensity to the median of the 10 preceding and 10 following exposures at the same pixel position and we replace outliers greater than $4~\sigma$ with its median value. The fraction of pixels we correct is less than 0.06\%. The centroid position of the stellar PSF is determined using a DAOPHOT-type Photometry Procedure, \texttt{GCNTRD}, from the IDL Astronomy Library\footnote{{\tt http://idlastro.gsfc.nasa.gov/homepage.html}}. We use the \texttt{APER} routine to perform aperture photometry with a circular aperture of variable radius, using radii of $1.5$ to $8$ pixels, in $0.5$ steps. The propagated uncertainties are derived as a function of the aperture radius; we adopt the one which provides the smallest errors. We find that the transit depths and errors vary only weakly with the aperture radius for all the light-curves analyzed in this project. The optimal aperture is found to have a radius of $4.0$~pixels. We estimate the background by fitting a Gaussian to the central region of the histogram of counts from the full array. The contribution of the background to the total flux from the stars is low for both observations, from 0.1\% to 0.55\% depending on the images. Therefore, photometric errors are not dominated by fluctuations in the background. We used a sliding median filter to select and trim outliers in flux and position greater than $5~\sigma$. We also discarded the first half-hour of observations, which are affected by a significant telescope jitter before stabilization. The final number of photometric measurements used is $1570$. The raw time series is presented in the top panel of Figure~\ref{fig:spitzerlightcurves}. We find that the point-to-point scatter in the photometry gives a typical signal-to-noise ratio of $330$ per image, which corresponds to 92\% of the theoretical signal-to-noise.
Therefore, the noise is dominated by Poisson photon noise.

\section{Analysis}
\label{sec:analysis}

\subsection{Centroid shifts}
\label{sec:centroid}
We use a comparison of the photometric centroid in- and out-of-transit to determine which component contains the transit event.  These centroids have been measured for quarters 1-6 using two methods: a) a fit of the transit model to the whitened row and column centroid time series, which provides an average offset in row and column for each quarter, and b) centroiding of quarterly average in- and out-of-transit images, where the in-transit average is constructed from all in-transit observations in a quarter and the out-of-transit average is constructed from placing the same number of points on each side of each transit event.

Both methods measured essentially identical centroid offsets.  These offsets were used to reconstruct the position on the sky of the transiting object, using the methods described in appendix A of \citet{Jenkins2010}.  The final reconstructed transit source location is then the average of the reconstructed transit position over all quarters.  The distance of this average reconstructed position from component A is $0.025\pm0.024\,\arcsec$ (1.04 sigma) and from component B is $0.251\pm0.030\,\arcsec$ (8.33 sigma). We conclude that the transiting object is component A.  

\subsection{Spectroscopic parameters of the host star}
\label{sec:specParam}
As noted in Section \ref{sec:spectroscopy}, we cannot separate the two stellar components on the fiber of the spectrograph and we thus observed the light from both stars in the spectra. As argued in Section \ref{sec:spectroscopy}, we assumed that the two stars are physically associated and that they formed together at the same time. Since the stars have nearly the same temperature due to their position on the H-R diagram, we concluded that the small magnitude difference would result in an insignificant change in the host star parameters (see Section \ref{sec:propertiesHostStar} for details).

We derived stellar atmospheric parameters from both the HIRES template spectrum and the high SNR FIES spectra used for the orbit determination, which can all be used because they are not contaminated by absorption from an iodine cell.

For the HIRES spectrum, we used an analysis package known as Spectroscopy Made Easy \citep[SME;][]{Valenti1996}, along with the atomic line database of \cite{Valenti2005}. From the HIRES spectrum using SME, we found the following parameters: effective temperature $\teffstar=\kepcurSMEteff$\,K, metallicity $\feh=\kepcurSMEfeh$\,dex, projected rotational velocity $\vsini=\kepcurSMEvsin\,\kms$, and stellar surface gravity $\loggstar=\kepcurSMElogg$\,(cgs).

For the FIES spectra, we derived stellar parameters following \citet{Torres2002} and \citet{Buchhave2010}, and in addition we employ a new fitting scheme which is currently still under development and being readied for publication, allowing us to extract more precise stellar parameters from the high SNR FIES spectra. We mention these values here as a check on the SME values, but adopt the SME values because our tools are still under development. From the FIES spectra, we found  effective temperature $\teffstar=\kepcurFIESteff$\,K, metallicity $\feh=\kepcurFIESfeh$\,dex, projected rotational velocity $\vsini=\kepcurFIESvsin\,\kms$, and stellar surface gravity $\loggstar=\kepcurFIESlogg$\,(cgs). All values are in good agreement with the values derived from the SME analysis, within the uncertainties, except for the value of \vsini.

\subsection{Properties of the host star}
\label{sec:propertiesHostStar}
Global properties of the star including the mass and radius were determined with the help of the stellar evolution models from the series by \cite{Girardi2000}. Isochrones for a wide range of ages were compared against the effective temperature and metallicity from the Keck/HIRES spectra, and the mean stellar density, $\rho_{\star}$, as an indicator of luminosity. If we assume a circular orbit, then the mean stellar density is closely related to the normalized semimajor axis $a/R_{\star}$ \citep[see, e.g.,][]{Seager2003, Sozzetti2007}, which is one of the parameters solved for in the light curve solutions described below in Section~\ref{sec:lightCurveAnalysis}, and is often more accurate than the spectroscopic $\log g$.  In practice we used $a/R_{\star}$ rather than $\rho_{\star}$, and the comparison with the isochrones was coupled with the light curve solutions, which were carried out using the Markov Chain Monte Carlo technique.  Specifically, we derived a \emph{distribution} of stellar properties by comparing the isochrones with each value in the $a/R_{\star}$ chains paired with values for the temperature and metallicity drawn from Gaussian distributions centered on the spectroscopically determined values and their errors.

The presence of the visual companion detected in our high-resolution imaging adds a complication, as the extra flux reduces the depth of the transit and affects its overall shape in subtle ways, biasing the $a/R_{\star}$ parameter. The impact of this extra dilution depends on the magnitude difference of the companion in the Kepler band ($\Delta K\!p$), which we expect to be close to (but not necessarily the same as) the measured magnitude differences in other passbands ($\Delta V$, $\Delta R$, $\Delta I$, $\Delta J$ and $\Delta K_s$). We therefore proceeded by iteration, in parallel with the light curve solutions.
We initially ignored the dilution effect on $a/R_{\star}$, and inferred the absolute magnitude of the target in the $K\!p$ band from the best-fit isochrone. Assuming the companion is physically associated and the two stars share the same isochrone, we then determined its mass along the isochrone with the condition that the magnitude difference in $V$ be exactly equal to the measured value. We then read off the $\Delta K\!p$ value directly from the isochrone. We repeated this using each of the other magnitude difference measurements (taking those in $J$ and $K_s$ from the MMT and Palomar to be independent), and we averaged the resulting seven values of $\Delta K\!p$ to obtain $0.45 \pm 0.10$~mag. With the corresponding relative flux $F_{\rm B}/F_{\rm A}$ a new light curve solution was carried out, leading to an improved $a/R_{\star}$ distribution. This, in turn, was compared once again with the isochrones, and led to a slightly revised brightness difference of $\Delta K\!p = \kepMagDiff$~mag. A further iteration did not change this significantly.

As described in Section~\ref{sec:specParam}, we have determined the host star parameters from the composite spectra of the primary star A and the fainter companion B, since it is not possible to separate the two stars on the fiber/slit of the spectrographs. We estimate that the adopted magnitude difference of $\Delta K\!p =\kepMagDiff$~mag does not significantly affect the derived spectroscopic stellar properties of the host star A. The companion star B is estimated to be only 30~K hotter and have a stronger surface gravity of 0.15 dex compared to the host star. We therefore choose to ignore the effect of the dilution on the stellar parameters of the host star.

The resulting properties of the host star are listed in Table~\ref{tab:parameters}, in which the values correspond to the mode of the distributions and the uncertainties reported are the 68.3\% (1$\sigma$) confidence limits defined by the 15.8\% and 84.2\% percentiles in the cumulative distributions.

We estimated the distance to \kep\ based on isochrones by comparing against the measured magnitudes from the Kepler Input Catalog \citep{Brown2011}. We fitted the spectral energy distribution with magnitudes for the two stars taken from the Girardi isochrones  resulting in a  distance estimate of 980 pc. For an average angular separation of \kepSeparation\ the semi-major axis of the visual pair is approximately 280 AU, and with mass estimates of $1.51~\msun$ and $1.39~\msun$ for the two stars, the corresponding period is of the order of 2800 years.

\subsection{Light curve analysis}
\label{sec:lightCurveAnalysis}

We modeled the folded transit light curve assuming spherical star and planet having radius ratio $R_p/R_\star$. The second star adds its light to the total light curve with the observed flux ratio between stars B and A being $F_B/F_A$.  The planet was constrained to a circular Keplerian orbit parameterized by a period $P$, a normalized semi-major axis distance $a/R_\star$ and an inclination to the sky plane $i$.

The normalized transit light curve, $f(t)$, was calculated to be \begin{eqnarray}
	f(t) & = & 1-\lambda\left[z(t)/R_\star,\frac{R_p}{R_\star}, u_1,
u_2\right]/\left(1+\frac{F_B}{F_A}\right)
\end{eqnarray}
where $z(t)$ is the sky-projected separation of the centers of the star and planet and $\lambda$ is the fraction of the stellar disk blocked by the planet, given analytically by \citet{Mandel2002}.
The limb darkening coefficients $u_1$ and $u_2$ parameterize the radial brightness profile, $I(r)$, of a star as \begin{eqnarray}
	\frac{I(r)}{I(0)} & =& 1-u_1 \left(1-\sqrt{1-r^2}\right)-u_2 \left(1-\sqrt{1-r^2}\right)^2.
\end{eqnarray}

The continuously defined model, $f(t)$, was numerically integrated before being compared with the long cadence {\em Kepler} light curve.
In detail, for each measured time, $t_j$, we take $n_j$ uniform samples $t_{j,k}= t_j+k \Delta t_j-\tau_{\rm int}/2$, separated by $\Delta t_j = \tau_{\rm int}/n_j$, over the long cadence integration interval of $\tau_{\rm int} = 29.4$ minutes.  The flux at $t_j$ was found by computing the Gaussian quadrature of the continuous model fluxes $f(t_{j,k})$.  In practice, we took $n_j = 20$ for all times.

We determined the best-fit model to the data by minimizing the $\chi^2$ goodness-of-fit statistic including Gaussian penalties to restrict the flux ratio $F_B/F_A$, $e \cos \omega$ and $e \sin \omega$ to agree with the observed constraints:
\begin{eqnarray}
	\lefteqn{\chi^2 = \sum_{s} \frac{\left({\mathcal F}_s-F_s\right)^2}{\sigma^2}+\frac{\left(F_B/F_A-0.667\right)^2}{0.061^2}+ {} }
	\nonumber\\
	& & {}
	\frac{\left(e
	\sin \omega-0.035\right)^2}{0.017^2}+\frac{\left(e \cos \omega-0.0006\right)^2}{0.0099^2} 
\label{eq:chi2} 
\end{eqnarray}
where $F_s$ is the measured flux at time $t_s$ and $\sigma$ is the expected statistical error in the flux measurements.  We selected $\sigma = 82$~ppm such that the reduced-$\chi^2$ was unity for the best fit solution.

We determined the posterior probability distribution for the fitted parameters by using a Differential Evolution Markov chain Monte Carlo
(DE-MCMC) algorithm \citep{TerBraak2006} with a Metropolis-Hastings jump condition and a jump acceptance probability conditional on the likelihood $\mathcal{L'} \propto \exp(-\chi^2/2)$.  We computed chains for a parallel population of 90 members through $\approx$2 million generations requiring that approximately $25\%$ of jumps were accepted on average amongst all population members. The chains were checked
for adequate mixing and convergence by visual inspection and by observing that the number of links was much larger than the autocorrelation length (equal to the number of links at which the chain autocorrelation drops below one half) for any selected parameter.  We report the $15.8\%$ and $84.2\%$ values of the cumulative distribution for each parameter, marginalizing over the remaining parameters.

\subsection{Analysis of the Warm-Spitzer light curves}

We used a transit light curve model multiplied by instrumental decorrelation functions to measure the transit parameters and their uncertainties from the \wspitzer\ data as described in \cite{Desert2011}. We computed the transit light curves with the IDL transit routine \texttt{OCCULTNL} from \cite{Mandel2002}.
This function depends on one parameter: the planet-to-star radius ratio $R_p / R_\star$. The orbital semi-major axis to stellar radius ratio (system scale) $a / R_\star$, the impact parameter $b$, and the time of mid transit $T_c$ were fixed to the values derived from the \kepler\ light curve and corrected for the dilution (See Table \ref{tab:parameters}). We assumed that the limb-darkening is well-approximated by a non-linear law at infrared wavelengths with four coefficients \citep{Claret2000} that we set to their values computed by \cite{Sing2010}.

The \spitzer/IRAC photometry is known to be systematically affected by the
so-called \textit{pixel-phase effect} (see e.g., \citealt{Charbonneau2005,Knutson2008}). This effect is seen as oscillations in the measured fluxes with a period of approximately 70~min (period of the telescope pointing jitter) and an amplitude of approximately $2\%$ peak-to-peak. We decorrelated our signal in each channel using a linear function of time for the baseline (two parameters) and a quadratic function of the PSF position (four parameters) to correct the data for each channel. We performed a simultaneous Levenberg-Marquardt least-squares fit \citep{Markwardt2009} to the data to determine the transit and instrumental model parameters (7 in total). The errors on each photometric point were assumed to be identical, and were set to the $RMS$ of the residuals of the initial best fit obtained. To obtain an estimate of the correlated and systematic errors \citep{Pont2006} in our measurements, we used the residual permutation bootstrap, or ``Prayer Bead'', method as described in \citet{Desert2009}. In this method, the residuals of the initial fit are shifted systematically and sequentially by one frame, and then added to the transit light curve model before fitting again. We allowed asymmetric error bars spanning $34\%$ of the points above and below the median of the distributions to derive the $1~\sigma$ uncertainties for each parameter as described in \citet{Desert2011b}.

\subsection{Interpretation of the Warm-Spitzer observations}\label{results}
We compute the theoretical dilution factor by extrapolating the $K_s$-band measurements to the \spitzer\ bandpass at 4.5~\micron. We estimate that 36\% of the photons recorded during the observation come from the companion star. We conclude that the presence of the contaminating star decreases the effective transit depth of \kep\ by a factor $0.61$. We measure the transit depth (limb-darkening removed) of \kep\ at 4.5~\micron\ and find $1722^{+127}_{-138}$~ppm uncorrected for the dilution. This corresponds to $R_p / R_\star=0.0415^{+0.0015}_{-0.0017}$. Applying the dilution correction we find $R_p / R_\star=0.0531^{+0.0019}_{-0.0021}$ which is consistent with the value derived from the \kepler\ photometry at better than the 2-$\sigma$ level. Our \spitzer\ observations provide an independent confirmation that the transit signal is achromatic, which supports the planetary nature of \kepb.

\subsection{Dilution effect on the radial velocities}
\label{sec:RV_dilution}

\begin{figure}
\centering
\plotone{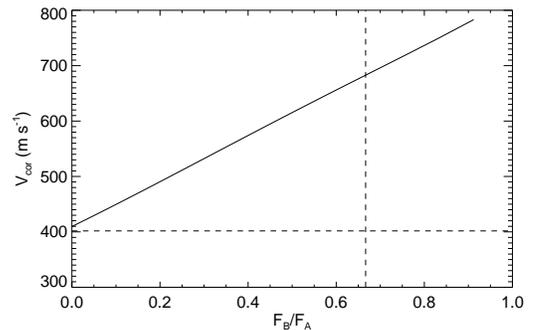}
\caption{The effect of the dilution of \kep\ on the measured radial velocities as a function of flux ratio between the companion star and the host star. The horizontal dashed line represents the observed semi-amplitude of the system and the vertical dashed line represents the adopted flux ratio of the two stars. If the stars have the same brightness, the corrected radial velocity is about twice the observed and at large magnitude differences, the corrected velocities approach the observed, as expected.}
\label{fig:RV_dilution}
\end{figure}

The measured radial velocities of the host star (A) are affected by the light contributed by the companion star (B), because the spectrum of B is assumed to be stationary in velocity, while the spectrum of A is Doppler shifted due to the gravitational pull of the planet. The amplitude of the observed radial velocities will thus be smaller than if the light from A had not been diluted, since the peak of the cross correlation function (CCF) from which we derive the radial velocities will be pulled toward the stationary CCF of B.

In order to assess the dilution effect on the radial velocities and thus the semi-amplitude of the orbit, we modeled the effect using the observed spectrum of \kep. We shifted the observed spectrum in $50~\ms$ increments and co-added the shifted spectrum, representing star A, with the same observed spectrum divided by a constant to simulate the stationary companion B at different flux ratios of $f = F_B/F_A$. We analyzed this composite spectrum using the same tools used to extract the radial velocities for the orbit. 

The relation between the artificially induced radial velocity shifts, $V_{in}$, and the resulting "measured" radial velocity shifts of the composite spectrum, $V_{out}$, is linear at given flux ratio, as expected: $V_{out} = a_f V_{in}$, where $a_f$ is the slope at a given flux ratio. $V_{in}$ thus represents the true (corrected) radial velocities of the host star ($V_{cor}$) and $V_{out}$ represents the observed radial velocities of the host star ($V_{obs}$). 

We carried out this analysis at different flux ratios, fitting the linear relation between $V_{in}$ and $V_{out}$, thus obtaining the slope $a_f$ at each flux ratio. We then fitted the slopes, $a_f$, themselves as a function of flux ratio with a 3rd order polynomial. This enables us to calculate the dilution effect for the system at any flux ratio:
\begin{eqnarray}
	V_{cor} = \frac{V_{obs}}{a_f} = \frac{V_{obs}}{c_0 + c_1f+c_2f^2+c_3f^3}
\end{eqnarray}
where $f$ is the flux ratio of the two stars and $c_i$ are the polynomial coefficients of the 3rd order polynomial. The dilution effect on the corrected radial velocities as a function of flux ratio can be seen in Figure \ref{fig:RV_dilution}. The horizontal dotted line represents the observed semi-amplitude of the system and the vertical dotted line represents the adopted flux ratio of the two stars. As a sanity check, we see that if the host star and companion have similar brightness, the corrected radial velocity is about twice the observed radial velocity and at large magnitude differences, the corrected radial velocity approaches the observed radial velocity, as expected.

The observed orbital semi-amplitude of \kep\ is $K_{obs} = \kepcurRVK\,\ms$. Since the two stars are nearly the same temperature, the dilution changes only minutely as a function of wavelength. We thus used the magnitude difference of $\Delta Kp = \kepMagDiff$ in all orders, and found the corrected semi-amplitude of the orbit to be $K_{cor} = \kepcurRVKcorr\,\ms$.

\subsection{Dilution effect on the planetary parameters}

The dilution of the nearly equal magnitude stellar companion significantly affects the derived planetary parameters of \kepb. The contamination affects the observed transit light curve depth and therefore the inferred radius ratio.  In addition, this dilution has a significant effect on the light curve profile affecting the inferred geometric orbital parameters, most notably the normalized semi-major axis, $a/R_\star$. If dilution effects are neglected, the mean stellar density estimate -- which is acutely sensitive to $a/R_\star$ -- used in conjunction with spectroscopic stellar constraints will yield significantly inaccurate derived stellar properties.

If we assume that the flux contribution from $B$ is zero (i.e., $F_B/F_A = 0$ and $K_{obs} = \kepcurRVK\,\ms$), we find that $R_{p,nocorr} = \kepcurPPrNoCorr\,\rjup$. Using the derived magnitude difference $\Delta Kp = \kepMagDiff$, however, we find the planetary radius to be $R_{p} = \kepcurPPr\,\rjup$, which is almost 10\% larger.

As described in Section~\ref{sec:RV_dilution}, the orbital semi-amplitude is also significantly affected by the dilution. Using the observed orbital semi-amplitude of $K_{obs} = \kepcurRVK\,\ms$, the uncorrected mass of \kepb\ is $M_{p,nocorr}=\kepcurPPmNoCorr\,\mjup$. After correction for dilution, the semi-amplitude increases to $K_{corr} = \kepcurRVKcorr\,\ms$, which in turn leads to a planetary mass that is significantly larger than before (by
$\sim$60\%): $M_{p} = \kepcurPPm\,\mjup$.

The effect of the dilution is much greater on the mass than on the radius of the transiting planet. As described above, the dilution of the observed transit light curve changes not only the depth of the transit, but also the light curve profile which in turn affects the inferred stellar density estimate. The radius of the planet is thus not affected greatly by the dilution, because these two effects work against each other. The stellar mass, however, is not strongly affected by the dilution and the effect on the planetary mass therefore comes almost entirely from correction of the orbital semi-amplitude.

\begin{deluxetable*}{lcc}
\tabletypesize{\scriptsize}
\tablecaption{System Parameters for \kep
\label{tab:parameters}}
\tablehead{
\colhead{Parameter}						& 
\colhead{Uncorrected} 				&
\colhead{Corrected}						\\
\colhead{ }						& 
\colhead{ } 				&
\colhead{(Adopted)}						\\
}

\startdata
\sidehead{\bf Transit and orbital parameters}
Orbital period $P$ (d)\tablenotemark{a,h}														& \dotfill & \kepcurLCP 	\\
Midtransit time $E$ (HJD)\tablenotemark{a}												&  \dotfill & \kepcurLCT 	\\
Transit duration (days)\tablenotemark{a,g}													& \kepcurDurationNoCorr & \kepcurDuration 	\\
Scaled semimajor axis $a/\rstar$\tablenotemark{a,b}								& \kepcurLCarNoCorr & \kepcurLCar	\\
Scaled planet radius \rpl/\rstar\tablenotemark{a,b}								& \kepcurLCrprstarNoCorr & \kepcurLCrprstar	\\
Impact parameter $b \equiv a \cos{i}/\rstar$\tablenotemark{a,b}		& \kepcurLCimpNoCorr & \kepcurLCimp	\\
Orbital inclination $i$ (deg)\tablenotemark{a,b}									& \kepcurLCiNoCorr & \kepcurLCi 	\\
Orbital semi-amplitude $K$ (\ms)\tablenotemark{b,c}								& \kepcurRVK & \kepcurRVKcorr	\\
$e \sin\omega$\tablenotemark{a,c}												&  \dotfill & $0.0350 \pm 0.0170$	\\
$e \cos\omega$\tablenotemark{a,c}												&  \dotfill & $0.0006 \pm 0.0099$	\\
Center-of-mass velocity $\gamma$ (\kms)\tablenotemark{c}					&   \dotfill & \kepcurRVmean	\\
\sidehead{\bf Observed stellar parameters}
Kepler magnitude $Kp$ \tablenotemark{f} & \dotfill & \kepMag \\
Effective temperature \teff\ (K)\tablenotemark{d}									&  \dotfill & \kepcurSMEteff	\\
Spectroscopic gravity \logg\ (cgs)\tablenotemark{d}								&  \dotfill & \kepcurSMElogg	\\
Metallicity \feh\tablenotemark{d}																	&  \dotfill & \kepcurSMEfeh	\\
Projected rotation \vsini\ (\kms)\tablenotemark{d}								&  \dotfill & \kepcurSMEvsin	\\
\sidehead{\bf Derived stellar parameters}
Mass \mstar (\msun)\tablenotemark{d,e}														& \kepcurYYmlongNoCorr & \kepcurYYmlong	\\
Radius \rstar (\rsun)\tablenotemark{d,e}	  											& \kepcurYYrlongNoCorr & \kepcurYYrlong	\\
Surface gravity \loggstar\ (cgs)\tablenotemark{d,e}								& \kepcurYYloggNoCorr & \kepcurYYlogg	\\
Luminosity \lstar\ (\lsun)\tablenotemark{d,e}											& \kepcurYYlumNoCorr & \kepcurYYlum	\\
Age (Gyr)\tablenotemark{d,e}																			& \kepcurYYageNoCorr & \kepcurYYage	\\
Distance (pc)\tablenotemark{e,f}																										& \dotfill & \kepcurXdist	\\
\sidehead{\bf Planetary parameters}
Mass \mpl\ (\mjup)\tablenotemark{a,b,c,d,e}												& \kepcurPPmNoCorr & \kepcurPPm	\\
Radius \rpl\ (\rjup, equatorial)\tablenotemark{a,b,c,d,e}					& \kepcurPPrNoCorr & \kepcurPPr	\\
Density \rhopl\ (\gcmc)\tablenotemark{a,b,c,d,e}									& \kepcurPPrhoNoCorr & \kepcurPPrho	\\
[-1.5ex]
\enddata
\tablenotetext{a}{
Based on the Kepler photometry.
}
\tablenotetext{b}{
Based on the dilution by the companion star.
}
\tablenotetext{c}{
Based on the FIES radial velocities.
}
\tablenotetext{d}{
Based on an SME analysis on the HIRES spectra.
}
\tablenotetext{e}{
Based on the Girardi stellar evolution models.
}
\tablenotetext{f}{
Based on the Kepler Input Catalog.
}
\tablenotetext{g}{
First to fourth contact point.
}
\tablenotetext{h}{
The actual orbital period differs fractionally from this value by $2.2 \pm 0.1 \times 10^{-5}$ as a result of time dilation for the quoted gamma velocity.
}
\end{deluxetable*}

\section{Discussion}
\label{sec:discussion}

\begin{figure}
\centering
\plotone{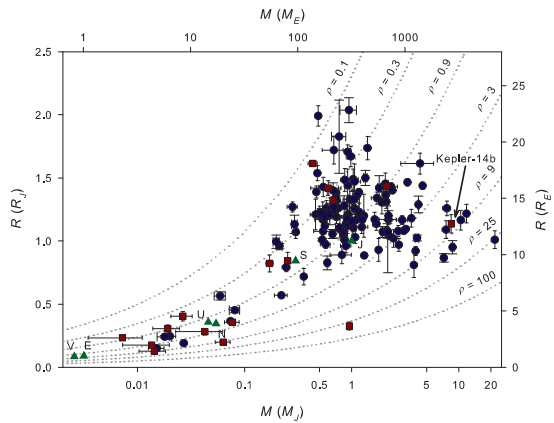}
\caption{
Mass--radius diagram  of currently  known transiting exoplanets. Kepler planets are shown as red squares and planets from other surveys are shown as blue circles. The Solar system planets are shown as green triangles. The dotted lines are isodensity curves (in \gcmc). \kep\ is one of the most massive transiting exoplanets discovered.  
}
\label{fig:MR_dia}
\end{figure}

We present the discovery of a transiting hot-Jupiter in a close visual binary. Had the visual companion not been detected, the planetary parameters for \kepb\ would have been significantly biased. The dilution ($\Delta Kp = \kepMagDiff$ in the Kepler band) results in a planetary mass that, if left uncorrected, is only 60\% of the correct value, and a planetary radius that is too small by about 10\%.

The close angular separation of this physically associated visual companion makes it essentially undetectable spectroscopically: the similar radial velocity as the main star means the spectrum is effectively single-lined, and the wide orbit ($P \sim 2800$ yr) implies motion that is slow enough that there are no measurable changes in the velocity of the primary due to this companion. It is only with high-resolution imaging that we were able to detect it. Many of the over 120 published transiting planets and the over 500 published radial velocity planets have not been subjected to high-resolution imaging. It is thus possible that some of the published exoplanets have incorrectly determined planetary parameters, if they have a stellar companion like \kep\ and the companion has not been taken into account. Since many of the published transiting planets have bright host stars, a campaign to gather high-resolution imaging of the host stars could be carried out with a modest amount of telescope time.

In this paper we confirm and characterize the planetary nature of \kepb, also known as \koi\ in \citet{Borucki2011}. \kepb\ has a period of $P = \kepcurLCP~\rm{days}$, a mass of $\mpl = \kepcurPPm\,\mjup$ and a radius of $\rpl = \kepcurPPr\,\rjup$, yielding a mean density of $\rhopl = \kepcurPPrho\,\gcmc$. Not taking the dilution into account, the derived mass and radius of the planet would be $M_{p,nocorr}=\kepcurPPmNoCorr\,\mjup$ and $R_{p,nocorr} = \kepcurPPrNoCorr\,\rjup$. 

\kepb\ is plotted on a mass--radius diagram in Figure \ref{fig:MR_dia}, which shows all the known transiting exoplanets. \kepb\ is one of the most massive transiting exoplanets discovered and is situated in a less dense part of the mass--radius diagram together with six other planets of similar mass. 

\acknowledgments Acknowledgments.
The work of L.A.B. was supported by the Carlsberg Foundation. Funding for this Discovery Mission is provided by NASA's Science Mission Directorate. This paper uses observations obtained with the Nordic Optical Telescope, operated on the island of La Palma jointly by Denmark, Finland, Iceland, Norway, and Sweden, in the Spanish Observatorio del Roque de los Muchachos of the Instituto de Astrofisica de Canarias. This work is also based on observations made with the Spitzer Space Telescope which is operated by the Jet Propulsion Laboratory, California Institute of Technology under a contract with NASA. Support for this work was also provided by NASA through an award issued by JPL/Caltech. 

{\it Facilities:} \facility{The Kepler Mission}, \facility{NOT (FIES)}, \facility{Keck:I (HIRES)},  \facility{Spitzer Space Telescope}, 
\facility{WIYN (Speckle)}, \facility{Palomar (AO)}, \facility{ARIES (AO)}

\bibliography{KOI-98_v04.bbl}

\end{document}